\documentclass[fleqn,12pt]{article}
\usepackage[affil-it]{authblk}
\usepackage[english]{babel}
\usepackage{blindtext}

\providecommand{\keywords}[1]{\textbf{\textit{Keywords:}} #1}
\providecommand{\msc}[1]{\textbf{\textit{2010 MSC:}} #1}

\usepackage{graphicx}
\usepackage{amsmath}
\usepackage{amsthm}
\usepackage{amssymb}
\usepackage{amsfonts}
\usepackage[hidelinks]{hyperref}
\usepackage{booktabs}

\usepackage{longtable}
\usepackage{cite}
\usepackage{mathrsfs}

\usepackage[title,titletoc,toc]{appendix}

\newtheorem{rem}{Remark}

\newtheorem{case}{Case}
\newtheorem{subcase}{Subcase}

\usepackage{chngcntr}
\counterwithin{case}{section}
\counterwithin{subcase}{case}
\counterwithin{exmp}{section}
\counterwithin{rem}{section}
\counterwithin{red}{section}
\counterwithin{prof}{section}

\usepackage[margin=1in,left=1in,right=1in]{geometry}
\usepackage{subcaption}
\usepackage{float}
\usepackage{placeins}
\allowdisplaybreaks
\usepackage{appendix}
\title{A revisit of symmetry analysis and group classifications of Boiti–Leon–Pempinelli system in (2+1)-dimensions
}

\author[1]{Manjit Singh\thanks{corresponding author: manjitcsir@gmail.com}}

\affil[1]{%
    Yadavindra College of Engineering Punjabi University Guru Kashi Campus Talwandi Sabo--151302, Punjab, India.}

\begin{document}

\maketitle
\begin{abstract}
In this paper, the Boiti–Leon–Pempinelli system in (2+1)-dimensions is revisited for Lie symmetries and invariant solutions. An infinite dimensional Lie algebra is obtained using the Lie invariance criterion and is further classified into one, two and three-dimensional optimal list of subalgebra. We obtain new explicit exact solutions involving arbitrary functions that have never been documented in previous work.
\end{abstract}
\\
\msc{70H07, 17B65, 17B67, 17B68.}\\
\section{Introduction}
The Lie symmetry analysis is  important in the study of partial differential equations for exact solutions and conservation laws \cite{anco,olverbook}. Much research has been conducted in the field of Lie symmetries, and with the development of powerful computational tools such as Maple and Mathematica, it is now possible to investigate partial differential equations for Lie symmetries with much ease. Over the last three decades, researchers have not left a single PDE for Lie symmetry analysis. However, where infinite-dimensional Lie algebra is concerned, there is still a lot of scope in Lie symmetry analysis. Because of the presence of arbitrary functions in the Lie algebra, Lie symmetry analysis for group classifications and invariant solutions becomes more difficult. The classic example is following Boiti–Leon–Pempinelli system (BLP system):
\begin{eqnarray}{\label{BLP:1}}
\begin{aligned}
L_{1}&\equiv u_{ty}-2\,u_{y}u_{x}-2\,uu_{xy}+u_{xxy}-2\,v_{xxx}=0,\\
L_{2}&\equiv v_{t}-2\,uv_{x}-v_{xx}=0,
\end{aligned}
\end{eqnarray}
where $u=u(x,y,t), v=v(x,y,t)$ and subscripts denote partial differentiation with respect to corresponding variables in the subscript. The equation \eqref{BLP:1} has been investigated for Lie symmetries in Refs. \cite{yu2010exact,kumar2014new,kumar2015some,zhao2017lie,krishnakumar2020lie,kumar2021some}. However, the symmetry analysis in all of these papers is incomplete and inadequate. In Ref. \cite{yu2010exact}, for example, an infinite Lie group involving arbitrary functions is obtained, and invariant solutions are constructed by fixing the arbitrary functions. In Refs. \cite{kumar2014new,kumar2015some}, the arbitrary functions in an infinite Lie group are again fixed and group invariant solutions are obtained, and the authors used the same technique to fix arbitrary functions in their recent work \cite{kumar2021some}. Furthermore, in Ref. \cite{zhao2017lie}, the authors made the same error of restricting arbitrary functions to polynomials. Similar deficiencies can be found in Ref. \cite{krishnakumar2020lie} in addition to sloppy group classifications. Moreover, in all these papers the authors have not emphasized on group classifications of Lie algebra of BLP system. Although in the Ref. \cite{zhao2017lie}, the authors have obtained group classification using Ibragimov's technique, but their Lie algebra is not general, rendering group classification incomplete.  And more importantly, the obvious physical symmetries can only be realized when arbitrary functions occurring in the infinite Lie group are first-order polynomials or exponential functions for some physical models. The symmetries which can be realized physically are limited to  translations, dilatation, rotations or Galilean boosts, and these types of physical symmetries can be obtained by restricting arbitrary functions to first-degree polynomials only. Therefore, randomly fixing arbitrary functions in the Lie algebra is completely incorrect and has no physical significance.
So in this work, the arbitrary functions in the infinite Lie group are kept intact and Lie symmetry analysis is performed without loss of generalization.

For Lie group of point transformations that leave system \eqref{BLP:1} invariant,  the following differential operator must be introduced
\begin{align}
    X=\xi^{x}\frac{\partial}{\partial x}+\xi^{y}\frac{\partial}{\partial y}+\xi^{t}\frac{\partial}{\partial t}+\eta^{u}\frac{\partial}{\partial u}+\eta^{v}\frac{\partial}{\partial v},
\end{align}
here each $\xi$ and $\eta$ depend on independent variables $x, y, t$ and dependent variables $u, v$. These are called infinitesimals that can be constructed from second order prolongation $X^{(2)}$,
\begin{align}
    X^{(2)}=X+\eta^{(1)u}_{i}\frac{\partial}{\partial u_{i}}+\eta^{(1)v}_{i}\frac{\partial}{\partial v_{i}}+\dots+\eta^{(2)u}_{i_{1}i_{2}}\frac{\partial}{\partial u_{i_{1}i_{2}}}+\eta^{(2)v}_{i_{1}i_{2}}\frac{\partial}{\partial v_{i_{1}i_{2}}}.
\end{align}
This prolonged operator will helps to extend all the derivatives appearing in the system \eqref{BLP:1} and the Lie invariance criterion require the system  to be invariant under this prolonged operator on the solution surface of the system \eqref{BLP:1}, that is,
\begin{align}
    \label{BLP:13}X^{(2)}L_{i}\big|_{L_{i}=0}=0, \quad i=1,2,
\end{align}
 and the most general solution invariance criterion \eqref{BLP:13} is obtained as follow:
\begin{eqnarray}{\label{BLP:2}}
\begin{aligned}
\xi^{x}=&2\,xf^{\prime}_{1}(t)+2\,f_{2}(t),\xi^{y}=h_{1}(y),\xi^{t}=4\,f_{1}(t),\\
\eta^{u}=&-2\,uf^{\prime}_{1}(t)-xf^{\prime\prime}_{1}(t)-f^{\prime}_{2}(t),\eta^{v}=-vh^{\prime}_{1}(t)+h_{2}(t),
\end{aligned}
\end{eqnarray}
where the arbitrary functions $f_{1}(t), f_{2}(t), h_{1}(y)$ and $h_{2}(y)$ will create following infinite-dimensional Lie algebra
 $\mathfrak{g}$:
\begin{eqnarray}{\label{BLP:3}}
\begin{aligned}
X_{1}=&2\,xf_{1}^{\prime}(t)\frac{\partial}{\partial x}+4\,f_{1}(t)\frac{\partial}{\partial t}-\left(2\,uf^{\prime}_{1}(t)+xf^{\prime\prime}_{1}(t)\right)\frac{\partial}{\partial u},\\
X_{2}=&2\,f_{2}(t)\frac{\partial}{\partial x}-f^{\prime}_{2}(t)\frac{\partial}{\partial u},\\
X_{3}=&h_{1}(y)\frac{\partial}{\partial y}-vh^{\prime}_{1}(y)\frac{\partial}{\partial v},\\
X_{4}=&h_{2}(y)\frac{\partial}{\partial v}.
\end{aligned}
\end{eqnarray}
The Lie algebra $\mathfrak{g}$ is solvable with following chain of ideals:
\begin{align*}
    \left\{X_{2}\right\}\subset\left\{X_{1},X_{2}\right\}\subset\left\{X_{1},X_{2},X_{4}\right\}\subset\left\{X_{1},X_{2},X_{3},X_{4}\right\},
\end{align*}
and it is closed under Lie commutation $[X_{i},X_{j}]=X_{i}X_{j}-X_{j}X_{i}$ and the Jacobi's identity $[X_{i},[X_{j},X_{k}]]+[X_{j},[X_{k},X_{i}]]+[X_{k},[X_{i},X_{j}]]=0$ is also satisfied. The results of all Lie commutations are listed in the \autoref{BLP_Table:1}.
\begin{table}[!ht]
\centering
    \begin{tabular}{c|cccc}
\hline\hline\\
  $[X_{i}, X_{j}]$ & $X_{1}$ & $X_{2}$ & $X_{3}$ & $X_{4}$  \\ \\\hline\hline\\
  $X_{1}$&0&$-X_{2}(f_{2}f^{\prime}_{1}-2\,f_{1}f^{\prime}_{2})$&0&0\\[1ex]
  $X_{2}$&$X_{2}(f_{2}f^{\prime}_{1}-2\,f_{1}f^{\prime}_{2})$&0&0&0\\[1ex]
  $X_{3}$&0&0&0&$X_{4}(h^{\prime}_{1}h_{2}+h_{1}h^{\prime}_{2})$\\[1ex]
  $X_{4}$&0&0&$-X_{4}(h^{\prime}_{1}h_{2}+h_{1}h^{\prime}_{2})$&0\\[1ex]
    \hline\hline
\end{tabular}
\caption{Commutation relations for Lie algebra \eqref{BLP:3}.}
\label{BLP_Table:1}
\end{table}
\begin{rem}
\normalfont It is worth noting that the Lie algebra obtained at \eqref{BLP:3} is more compact than the one obtained in Ref. \cite{zhao2017lie}. The determining equations do not appear to be solved judiciously in Ref. \cite{zhao2017lie}.
\end{rem}
\begin{rem}\label{BLP-Rem-1.2}
\normalfont A conclusion can be drawn from the structure of infinite-dimensional Lie algebra \eqref{BLP:3} that it contains Virasoro subalgebra with is typical characteristics of integrable systems \cite{david1985subalgebras,champagne1988infinite,faucher1993symmetry,paquin1990group}. It is therefore very likely that the Boiti–Leon–Pempinelli system \eqref{BLP:1} may be completely integrable, though it is not quite rigorous way to claim for the integrability.  
\end{rem}
\section{Multi-dimensional group classifications}
We define adjoint transformation
\begin{align}
    \label{BLP:4}\mathrm{Ad}_{\mathrm{exp}(\epsilon X_{i})}(X_{j})=\mathrm{e}^{-\epsilon X_{i}}X_{j}\mathrm{e}^{\epsilon X_{i}}=\tilde{X_{j}}(\epsilon).
\end{align}
 The adjoint transformation \eqref{BLP:4} can be written through Lie brackets using Campbell-Hausdorff formula as
\begin{align}
\label{BLP:5}\text{Ad}_{\exp(\epsilon X_{i})}\left(X_{j}\right) = X_{j}-\epsilon [X_{i},X_{j}]+\frac{\epsilon^{2}}{2}[X_{i},[X_{i},X_{j}]]-\dots,
\end{align}
where $[. , .]$ is Lie bracket defined by \autoref{BLP_Table:1}. The relation \eqref{BLP:5} helps to compile a table of the adjoint actions among each element in \eqref{BLP:3}. All such adjoint actions are listed in the \autoref{BLP_Table:2}.
\begin{table}[!ht]
\centering
    \begin{tabular}{c|cccc}
\hline\hline\\
  $\text{Ad}_{\exp(\epsilon X_{i})}\left(X_{j}\right)$ & $X_{1}$ & $X_{2}$ & $X_{3}$ & $X_{4}$  \\ \\\hline\hline\\
  $X_{1}$&$X_{1}$&$\mathrm{e}^{\epsilon}X_{2}$&$X_{3}$&$X_{4}$\\[1ex]
  $X_{2}$&$X_{1}-\epsilon X_{2}$&$X_{2}$&$X_{3}$&$X_{4}$\\[1ex]
  $X_{3}$&$X_{1}$&$X_{2}$&$X_{3}$&$\mathrm{e}^{-\epsilon}X_{4}$\\[1ex]
  $X_{4}$&$X_{1}$&$X_{2}$&$X_{3}+\epsilon X_{4}$&$X_{4}$\\[1ex]
    \hline\hline
\end{tabular}
\caption{Commutation relations for Lie algebra \eqref{BLP:3}}
\label{BLP_Table:2}
\end{table}
For general element $X=\sum_{i=1}^{4}$ in the Lie algebra \eqref{BLP:3}, the successive application of the formula \eqref{BLP:4} yields full adjoint action as follows:
\begin{align}
  \label{BLP:6} \text{Ad}_{\exp(\epsilon_{1} X_{1})}\text{Ad}_{\exp(\epsilon_{3} X_{3})}\text{Ad}_{\exp(\epsilon_{4} X_{4})}\text{Ad}_{\exp(\epsilon_{2} X_{2})}(X)=\sum_{i=1}^{4}\tilde{a}_{i}X_{i},
\end{align}
where the coefficient $\tilde{a}_{i}$ are given as follows:
\begin{eqnarray}{\label{BLP:7}}
\begin{aligned}
\tilde{a}_{1}=&a_{1}, \tilde{a}_{2}=\mathrm{e}^{\epsilon_{1}}(-a_{1}\epsilon_{2}+a_{2}),\\
\tilde{a}_{3}=&a_{3}, \tilde{a}_{4}=\mathrm{e}^{-\epsilon_{3}}(a_{3}\epsilon_{4}+a_{4}).
\end{aligned}
\end{eqnarray}
As described in the Ref. \cite{singh2020infinite}, the Killing form and the invariant function can be calculated as $K(X,X)=a_{1}^{2}+a_{3}^{2}$ and  $\phi=f(a_{1},a_{3})$ (for arbitrary function $f$) respectively. Both agree with the fact that $a_{1}$ and $a_{3}$ are invariants of full adjoint action \eqref{BLP:7}.
\subsection{One-dimensional optimal system}
The procedure for the constructing of an optimal system is quite wild, but the construction becomes simple and straightforward when invariants of full-adjoint are known. For example, in the present case, $a_{1}$ and $a_{3}$ are invariants of full-adjoint, by restricting their value the simplification of \eqref{BLP:6} becomes quite easy. The detailed simplification of \eqref{BLP:6} with the help of four different cases is described below.
\begin{case}
\normalfont When $a_{1}\neq 0, a_{3}\neq 0$,  we set $\epsilon_{1}=\epsilon_{3}=0$ (this means in \eqref{BLP:6}, the adjoint actions $\text{Ad}_{\exp(\epsilon_{1} X_{1})}, \text{Ad}_{\exp(\epsilon_{3} X_{3})}$  are being inactivated), and on setting $\epsilon_{2}=\frac{a_{2}}{a_{1}}$ and $\epsilon_{4}=-\frac{a_{4}}{a_{3}}$, the coefficients $\tilde{a}_{2}, \tilde{a}_{4}$ vanish. The general element $X$ is thus simplifies to $X_{1}+\alpha\,X_{3}$ for $\alpha=\frac{a_{3}}{a_{1}}$.
\end{case}
\begin{case}
\normalfont When $a_{1}= 0, a_{3}\neq 0$, on setting $a_{3}=1, \epsilon_{3}=0, \epsilon_{4}=-a_{4}$, the general element $X$ reduces to $a_{2}\mathrm{e}^{\epsilon_{1}}X_{2}+X_{3}$ and the coefficient of $X_{2}$ can be scaled to $\pm 1$ by taking $\epsilon_{1}=\log\left|\frac{1}{a_{1}}\right|$, and hence the final simplification would be $X_{3}\pm X_{2}$.
\begin{subcase}
\normalfont When $a_{1}= 0, a_{2}= 0, a_{3}\neq 0$, the obvious simplification of general element $X$ would be $X_{3}$.
\end{subcase}
\end{case}
\begin{case}
\normalfont When $a_{1}\neq 0, a_{3}= 0$, on setting $a_{1}=1, \epsilon_{1}=0, \epsilon_{2}=a_{2}$, the general element $X$ reduces to $X_{1}+a_{4}\mathrm{e}^{-\epsilon_{3}}X_{4}$ and the coefficient of $X_{4}$ can be scaled to $\pm 1$ by taking $\epsilon_{3}=\log|a_{4}|$, and hence the final simplification would be $X_{1}\pm X_{4}$.
\end{case}
\begin{case}
\normalfont When $a_{1}= 0, a_{3}= 0$, for $\epsilon_{1}=0$, the general element $X$ reduces to $a_{2}X_{2}+a_{4}\mathrm{e}^{-\epsilon_{3}}X_{4}$, and the final simplification can be achieved as $X_{2}\pm X_{4}$ by taking $\epsilon_{3}=\log\left|\frac{a_{4}}{a_{2}}\right|$.
\begin{subcase}
\normalfont When $a_{1}= 0, a_{2}=0, a_{3}= 0$, the obvious simplification of general element $X$ would be $X_{4}$.
\end{subcase}
\begin{subcase}
\normalfont When $a_{1}= 0, a_{3}=0, a_{4}= 0$, the obvious simplification of general element $X$ would be $X_{2}$.
\end{subcase}
\end{case}
The above cases and sub-cases can be summed up to one-dimensional optimal list sub-algebras $\Theta_{1}$ as follow:
\begin{eqnarray}{\label{BLP:8}}
\begin{aligned}
&X_{1}+\alpha\,X_{3}, X_{3}+\epsilon\,X_{2}, X_{3},\\
&X_{1}+\epsilon\,X_{4}, X_{2}+\epsilon\,X_{4}, X_{4}, X_{2}, \quad \alpha=\frac{a_{3}}{a_{1}}, \epsilon=\pm 1.
\end{aligned}
\end{eqnarray}
\subsection{Two-dimensional optimal system}{\label{BLP_sec2.2}}
All two-dimensional subalgebras are of type $\left(X_{i}, X\right)$, where $X_{i}$ runs through optimal list of subalgebras \eqref{BLP:8} and $X$ is general element from Lie algebra \eqref{BLP:3} such that
\begin{align}
    \label{BLP:9}\left[X_{i}, \sum_{i=1}^{4}a_{i}X_{i}\right]=\alpha\,X_{i},
\end{align}
where $\alpha$ is constant and the constants $a_{i}$ can be determined by equating coefficients of $X_{i}$ on the both sides of \eqref{BLP:9}. In addition, a detailed procedure for the construction of a two-dimensional optimal subalgebra list can be seen in Ref. \cite{hu2016constructing} and a recent work by the author himself \cite{singh2020infinite}. The following is the list of two-dimensional subalgebras $\Theta_{2}$ formed by this construction:
\begin{eqnarray}{\label{BLP:10}}
\begin{aligned}
&\mathcal{G}_{1}(X_{1}+\alpha\,X_{3},X_{2}),\\
    &\mathcal{G}_{2}(X_{1}+\alpha\,X_{3},X_{4}),\\
    &\mathcal{G}_{3}(X_{3}+\epsilon\,X_{2},X_{4}),\\
    &\mathcal{G}_{4}(X_{3},X_{1}),\\
    &\mathcal{G}_{5}(X_{1}+\epsilon\,X_{4},X_{2}),\\
    &\mathcal{G}_{6}(X_{2}+\epsilon\,X_{4},X_{1}+X_{3}),\\
    &\mathcal{G}_{7}(X_{4},X_{1}),\\
    &\mathcal{G}_{8}(X_{2},X_{3}),\\
    &\mathcal{G}_{9}(X_{2},X_{4}).
\end{aligned}
\end{eqnarray}

\subsection{Three-dimensional optimal system}
The three-dimensional optimal list of subalgebra $\Theta_{3}$ can be constructed in analogous manner as described in \autoref{BLP_sec2.2}. The following list of three-dimensional optimal subalgebras is obtained:
\begin{eqnarray}{\label{BLP:11}}
\begin{aligned}
&\mathcal{H}_{1}(X_{1},X_{3},X_{2}),\\
    &\mathcal{H}_{2}(X_{1},X_{4},X_{2}),\\
    &\mathcal{H}_{3}(X_{1},X_{3},X_{4}),\\
    &\mathcal{H}_{4}(X_{2},X_{3},X_{4}),\\
    &\mathcal{H}_{5}(X_{1}+X_{3},X_{2},X_{4}),\\
    &\mathcal{H}_{6}(X_{1}-X_{3},X_{2},X_{4}),\\
    &\mathcal{H}_{7}(X_{1}+\alpha\,X_{3},X_{2},X_{4}).
\end{aligned}
\end{eqnarray}
In following sections, we shall present similarity reductions corresponding to optimal list of subalgebras \eqref{BLP:8},  \eqref{BLP:10} and  \eqref{BLP:11}.
\section{Similarity reductions and invariant solutions}
The number of independent variables in BLP-equation \eqref{BLP:1} can be reduced by performing similarity reduction, the procedure is well described in the references \cite{ovsi,olverbook,anco}. As we shall see, the each subalgebra in the list \eqref{BLP:8} can reduce the number of variables  by one, each subalgebra in the list \eqref{BLP:10} can reduce the number of variables  by two, and each subalgebra in the list \eqref{BLP:11} can reduce the number of variables  by three. The reduction using three-dimensional subalgebra is very interesting as the final reduction shall be the analytical solution for the BLP-equation.
\subsection{Reductions under one-dimensional subalgebra}
In this section, the similarity reductions under one-dimensional optimal list of subalgebras \eqref{BLP:8} is given. To illustrate the reduction, we consider subalgebra $X_{1}+\alpha\,X_{3}$, and the following characteristics equations may be written:
\begin{align*}
    \frac{dx}{2\,xf^{\prime}_{1}(t)}=\frac{dy}{\alpha\,h_{1}(y)}=\frac{dt}{4\,f_{1}(t)}=\frac{du}{-(2\,uf^{\prime}_{1}(t)+xf^{\prime\prime}_{1}(t))}=\frac{dv}{-\alpha\,vh^{\prime}_{1}(y)}.
\end{align*}
On solving these characteristics equations the similarity variables and reduction fields may be quickly derived as follow:
\begin{eqnarray}{\label{BLP:12}}
\begin{aligned}
&\xi=\frac{x^{2}}{f_{1}(t)},\eta=\int\frac{1}{\alpha\,h_{1}(y)}dy-\int\frac{1}{4\,f_{1}(t)}dt,\\
&u=-\frac{xf^{\prime}_{1}(t)}{4\,f_{1}(t)}+\frac{1}{\sqrt{f_{1}(t)}}\cdot F(\xi,\eta), v=\frac{1}{h_{1}(y)}\cdot G(\xi,\eta).
\end{aligned}
\end{eqnarray}
Substituting the reduction fields $u$ and $v$ from \eqref{BLP:12} to BLP-equation \eqref{BLP:1} shall provide the following reduced system:
\begin{subequations}{\label{BLP:17}}
\begin{align}
&F_{\eta\eta}+16\xi^{\frac{1}{2}}F_{\xi}F_{\eta}+16\xi^{\frac{1}{2}}FF_{\xi\eta}-16\,\xi F_{\xi\xi\eta}-8F_{\xi\eta}+64\alpha\,\xi^{\frac{3}{2}}G_{\xi\xi\xi}+96\alpha\,\xi^{\frac{1}{2}}G_{\xi\xi}=0,\\
&G_{\eta}+16\,\xi G_{\xi\xi}+8\,G_{\xi}+16\,\xi^{\frac{1}{2}}FG_{\xi}=0.
\end{align}
\end{subequations}
The natural solution for the system \eqref{BLP:17} may be obtained as follows:
\begin{align*}
    F=-\frac{1}{2}\xi^{-\frac{1}{2}}, G=c_{1}\xi^{\frac{1}{2}}+c_{2}\xi-16\,\xi\eta+c_{3}.
\end{align*}
The final solution of the system \eqref{BLP:1} is obtained as follow:
\begin{subequations}\label{BLP:18}
\begin{align}
&u^{1}=-\frac{xf^{\prime}_{1}(t)}{4\,f_{1}(t)}-\frac{1}{2x},\\ &v^{1}=\frac{1}{h_{1}(y)}\left(\frac{c_{1}\,x}{\sqrt{f_{1}(t)}}+\frac{c_{2}\,x^{2}}{f_{1}(t)}-16\,\frac{x^{2}}{f_{1}(t)}\left(\int\frac{1}{\alpha\,h_{1}(y)}dy-\int\frac{1}{4\,f_{1}(t)}dt\right)+c_{3}\right).
\end{align}
\end{subequations}

\begin{table}[!ht]
\centering
    \begin{tabular}{l|l|l}
\hline\hline\\
  Subalgebra & Similarity Transformations & Reduced Equations  \\ \\\hline\hline
  $X_{2}$&$\begin{aligned}
  &\xi=y, \eta=t\\
  &u=-\frac{x f_{2}^{\prime}(t)}{f_{2}(t)}+F(\xi,\eta), v=G(\xi,\eta)
  \end{aligned}$&$\begin{aligned}
  &F_{\xi\eta}+\frac{1}{f_{2}(\eta)}\frac{d}{d\eta}(f_{2}(\eta))F_{\xi}=0, G_{\eta}=0
  \end{aligned}$\\[1ex]\hline
  $X_{3}$&$\begin{aligned}
  &\xi=x,\eta=t\\&u=F(\xi,\eta),v=\frac{1}{h_{1}(y)}G(\xi,\eta)
  \end{aligned}$&$\begin{aligned}
  G_{\xi\xi\xi}=0, G_{\eta}-2FG_{\xi}-G_{\xi\xi}=0
  \end{aligned}$\\[1ex]\hline
  $X_{3}+\epsilon\,X_{2}$&$\begin{aligned}
  &\xi=\frac{x}{2\epsilon f_{2}(t)}-\int\frac{1}{h_{1}(y)}dy, \eta=t\\
  &u=-\frac{xf^{\prime}_{2}(t)}{2f_{2}(t)}+F(\xi,\eta),\\ &v=\frac{1}{h_{1}(y)}G(\xi,\eta)
  \end{aligned}$&$\begin{aligned}
  &4\epsilon^{3}f_{2}^{3}(\eta)F_{\xi\eta}+4\epsilon^{3}f_{2}^{2}(\eta)f_{2}^{\prime}(\eta)F_{\xi}-4\epsilon^{2}f_{2}^{2}(\eta)F_{\xi}^{2}\\&-4\epsilon^{2}f_{2}^{2}(\eta)FF_{\xi\xi}+\epsilon f_{2}(\eta)F_{\xi\xi\xi}+G_{\xi\xi\xi}=0,\\
  &G_{\xi\xi}-4\epsilon^{2}f_{2}^{2}(\eta)G_{\eta}+4\epsilon f_{2}(\eta)FG_{\xi}=0
  \end{aligned}$\\[1ex]\hline
  $X_{1}+\epsilon\,X_{4}$&$\begin{aligned}
  &\xi=\frac{x^{2}}{f_{1}(t)}, \eta=y\\
  &u=-\frac{xf_{1}^{\prime}(t)}{4f_{1}(t)}+\frac{1}{\sqrt{f_{1}(t)}}F(\xi,\eta),\\
  &v=\frac{\epsilon h_{2}(y)}{4}\int\frac{1}{f_{1}(t)}dt+G(\xi,\eta)
  \end{aligned}$&$\begin{aligned}
  &4\xi^{\frac{1}{2}}FF_{\xi\eta}-4\xi F_{\xi\xi\eta}+4\xi^{\frac{1}{2}}F_{\xi}F_{\eta}+16\xi^{\frac{3}{2}}G_{\xi\xi\xi}\\&-2F_{\xi\eta}+24\xi^{\frac{1}{2}}G_{\xi\xi}=0,\\
  &16\xi^{\frac{1}{2}} FG_{\xi}-\epsilon h_{2}(\eta)+16\xi G_{\xi\xi}+8G_{\xi}=0
  \end{aligned}$\\[1ex]\hline
  $X_{2}+\epsilon\,X_{4}$&$\begin{aligned}
  &\xi=y,\eta=t,\\
  &u=-\frac{xf_{2}^{\prime}(t)}{2\,f_{2}(t)}+F(\xi,\eta),\\
  &v=\frac{\epsilon x h_{2}(y)}{2f_{2}(t)}+G(\xi,\eta)
  \end{aligned}$&$\begin{aligned}
  &F_{\xi\eta}+\frac{f_{2}^{\prime}(\eta)}{f_{2}(\eta)}F_{\xi}=0,\\
  &f_{2}(\eta)G_{\eta}-\epsilon h_{2}(\xi)F=0
  \end{aligned}$\\[1ex]
    \hline\hline
\end{tabular}
\caption{Similarity reductions corresponding to each subalgebra in the one-dimensional optimal list \eqref{BLP:8}.}
\label{BLP_Table:3}
\end{table}
The rest of the similarity reductions corresponding to remaining subalgebras in the list \eqref{BLP:8} are given  in the \autoref{BLP_Table:3}. The invariant solutions for remaining reduced systems are listed into following five cases:
\begin{case}
\normalfont Under $X_{2}$. The reduced system can be explicitly solved, and reduced fields are obtained as follow:
\begin{align*}
    F=\psi_{1}(\eta)+\frac{\phi_{1}(\xi)}{f_{2}(\eta)}, G=\phi_{2}(\xi).
\end{align*}
The final solution of the system \eqref{BLP:1} is obtained as follow:
\begin{subequations}\label{BLP:19}
\begin{align}
    u^{2}=&-\frac{x f_{2}^{\prime}(t)}{f_{2}(t)}+\psi_{1}(t)+\frac{\phi_{1}(y)}{f_{2}(t)},\\
    v^{2}=&\phi_{2}(y).
\end{align}
\end{subequations}
\end{case}
\begin{case}
\normalfont Under $X_{3}$. The reduced system can be explicitly solved, and reduced fields are obtained as follow:
\begin{align*}
    F=&{\frac { \left( {\frac {\rm d}{{\rm d}\eta}}\psi_{{1}} \left( \eta
 \right)  \right) {\xi}^{2}+2\, \left( {\frac {\rm d}{{\rm d}\eta}}
\psi_{{2}} \left( \eta \right)  \right) \xi+2\,{\frac {\rm d}{{\rm d}
\eta}}\psi_{{3}} \left( \eta \right) -2\,\psi_{{1}} \left( \eta
 \right) }{4\,\psi_{{1}} \left( \eta \right) \xi+4\,\psi_{{2}} \left( 
\eta \right) }},\\
G=&\frac{1}{2}\,\psi_{{1}} \left( \eta \right) {\xi}^{2}+\psi_{{2}} \left( \eta
 \right) \xi+\psi_{{3}} \left( \eta \right).
\end{align*}
The final solution of the system \eqref{BLP:1} is obtained as follow:
\begin{subequations}\label{BLP:20}
\begin{align}
    u^{3}=&{\frac { \left( {\frac {\rm d}{{\rm d}t}}\psi_{{1}} \left( t
 \right)  \right) {x}^{2}+2\, \left( {\frac {\rm d}{{\rm d}t}}\psi_{{2
}} \left( t \right)  \right) x+2\,{\frac {\rm d}{{\rm d}t}}\psi_{{3}}
 \left( t \right) -2\,\psi_{{1}} \left( t \right) }{4\,\psi_{{1}}
 \left( t \right) x+4\,\psi_{{2}} \left( t \right) }}
,\\
v^{3}=&\frac{1}{h_{1}(y)}\left(\frac{1}{2}\,\,\psi_{{1}} \left( t \right) {x}^{2}+\psi_{{2}} \left( t
 \right) x+\psi_{{3}} \left( t \right)
\right).
\end{align}
\end{subequations}
\end{case}
\begin{case}
\normalfont Under $X_{3}+\epsilon\,X_{2}$. The reduced system corresponding to this subalgebra can be further reduced using symmetries, yielding two sets of invariant solutions:
\begin{align*}
    \left\{F=\epsilon\,f_{2}(\eta)\frac{d}{d\eta}(\psi_{1}(\eta)),
    G=\xi+\psi_{1}(\eta)\right\},
\end{align*}
 \begin{align*}
    \left\{F=-\frac{\epsilon}{f_{2}(\eta)},
     G=c_{1}\left[\xi-\int\frac{1}{f_{2}^{2}(\eta)}d\eta\right]+c_{2}\right\}.
 \end{align*}
 The final solution of the system \eqref{BLP:1} is obtained as follow:
 \begin{subequations}\label{BLP:21}
 \begin{align}
     &u^{4}=-\frac{xf^{\prime}_{2}(t)}{2f_{2}(t)}+\epsilon\,f_{2}(t)\frac{d}{dt}(\psi_{1}(t)),\\ &v^{4}=\frac{1}{h_{1}(y)}\left(\frac{x}{2\epsilon f_{2}(t)}-\int\frac{1}{h_{1}(y)}dy+\psi_{1}(t)\right).
 \end{align}
 \end{subequations}
 \begin{subequations}\label{BLP:22}
 \begin{align}
     &u^{5}=-\frac{xf^{\prime}_{2}(t)}{2f_{2}(t)}-\frac{\epsilon}{f_{2}(t)},\\ &v^{5}=\frac{1}{h_{1}(y)}\left(c_{1}\left[\frac{x}{2\epsilon f_{2}(t)}-\int\frac{1}{h_{1}(y)}dy-\int\frac{1}{f_{2}^{2}(t)}dt\right]+c_{2}\right).
 \end{align}
 \end{subequations}
\end{case}

\begin{case}
\normalfont Under $X_{1}+\epsilon\,X_{4}$. The reduced system corresponding to  this subalgebra admits four sets of invariant solutions which are given below:
\begin{align*}
    \left\{F  =\frac{\epsilon\,h_{{2}}\left( 
\eta \right)}{16}  ,G  =2\,\sqrt {\xi}+\phi_{1}\left( 
\eta \right)\right\},
\end{align*}
\begin{align*}
    \left\{ F  ={\frac {\epsilon\,h_{{2}}
 \left(\eta \right) }{16\,\eta}},G  =2\,\sqrt {\xi
}\eta+\phi_{1}\left( \eta \right)  \right\},
\end{align*}
\begin{align*}
    \left\{ F  ={\frac {-8\,c_{{2}}+\epsilon}{16\,
c_{{2}}\sqrt {\xi}+8\,c_{{3}}}},G  =h_{{2}}
 \left( \eta \right)  \left( \sqrt {\xi}c_{{3}}+c_{{2}}\xi+c_{{1}}+
\eta \right)  \right\},
\end{align*}
\begin{align*}
    \left\{ F  ={\frac {-8\,c_{{2}}+\epsilon}{16\,
c_{{2}}\sqrt {\xi}+8\,c_{{3}}}},G  =\frac{h_{{2}
}\left( \eta \right)}{2}   \left( 2\,\sqrt {\xi}c_{{3}}+2\,c_{{2}}\xi+{
\eta}^{2}+2\,c_{{1}} \right)  \right\}. 
\end{align*}
The final solutions of the system \eqref{BLP:1} is obtained as follow:
\begin{subequations}\label{BLP:23}
\begin{align}
  &u^{6}=-\frac{xf_{1}^{\prime}(t)}{4f_{1}(t)}+\frac{\epsilon h_{2}(y)}{16\sqrt{f_{1}(t)}},\\
  &v^{6}=\frac{\epsilon h_{2}(y)}{4}\int\frac{1}{f_{1}(t)}dt+\frac{2x}{\sqrt{f_{1}(t)}}+\phi_{1}\left( 
y \right),
  \end{align}
\end{subequations}
\begin{subequations}\label{BLP:24}
\begin{align}
  &u^{7}=-\frac{xf_{1}^{\prime}(t)}{4f_{1}(t)}+\frac{\epsilon h_{2}(y)}{16y\sqrt{f_{1}(t)}},\\
  &v^{7}=\frac{\epsilon h_{2}(y)}{4}\int\frac{1}{f_{1}(t)}dt+\frac{2xy}{\sqrt{f_{1}(t)}}+\phi_{1}\left( 
y \right),
  \end{align}
\end{subequations}
\begin{subequations}\label{BLP:25}
\begin{align}
    &u^{8}=-\frac{xf_{1}^{\prime}(t)}{4f_{1}(t)}+{\frac {-8\,c_{{2}}+\epsilon}{16\,
c_{{2}}x+8\,c_{{3}}\sqrt{f_{1}(t)}}},\\
  &v^{8}=\frac{\epsilon h_{2}(y)}{4}\int\frac{1}{f_{1}(t)}dt+h_{{2}}
 \left(y\right)  \left( \frac{c_{{3}}x}{\sqrt{f_{1}(t)}}+\frac{c_{{2}}x^{2}}{f_{1}(t)}+c_{{1}}+
y \right),
\end{align}
\end{subequations}
\begin{subequations}\label{BLP:26}
\begin{align}
    &u^{9}=-\frac{xf_{1}^{\prime}(t)}{4f_{1}(t)}+{\frac {-8\,c_{{2}}+\epsilon}{16\,
c_{{2}}x+8\,c_{{3}}\sqrt{f_{1}(t)}}},\\
  &v^{9}=\frac{\epsilon h_{2}(y)}{4}\int\frac{1}{f_{1}(t)}dt+h_{{2}}
 \left(y\right)  \left( \frac{c_{{3}}x}{\sqrt{f_{1}(t)}}+\frac{c_{{2}}x^{2}}{f_{1}(t)}+c_{{1}}+
\frac{y^{2}}{2} \right),
\end{align}
\end{subequations}
\end{case}

\begin{case}
\normalfont Under $X_{2}+\epsilon\,X_{4}$. The reduced system can be explicitly solved, and reduced fields are obtained as follow:
\begin{align*}
    F=&\psi_{1}(\eta)+\frac{\phi_{1}(\xi)}{f_{2}(\eta)},\\
    G=&\epsilon h_{2}(\xi)\int\frac{\psi_{1}(\eta)f_{2}(\eta)+\phi_{1}(\xi)}{f_{2}^{2}(\eta)}d\eta+\phi_{2}(\xi).
\end{align*}
The final solution of the system \eqref{BLP:1} is obtained as follow:
\begin{subequations}\label{BLP:27}
\begin{align}
  &u^{10}=-\frac{xf_{2}^{\prime}(t)}{2\,f_{2}(t)}+\psi_{1}(t)+\frac{\phi_{1}(y)}{f_{2}(t)},\\
  &v^{10}=\frac{\epsilon x h_{2}(y)}{2f_{2}(t)}+\epsilon h_{2}(y)\int\frac{\psi_{1}(t)f_{2}(t)+\phi_{1}(y)}{f_{2}^{2}(t)}dt+\phi_{2}(y).
  \end{align}
\end{subequations}

\end{case}
\begin{rem}
\normalfont The solutions $\{u^{i},v^{i}\}, i=1\dots10$, given in  \eqref{BLP:18}--\eqref{BLP:27} are highly explicit as each of the solution includes several arbitrary functions, and could be used to describe physical phenomena with greater degrees of freedom. And we believe these solutions have never been documented before not even in the recent work on BLP system \cite{yu2010exact,kumar2014new,kumar2015some,zhao2017lie,krishnakumar2020lie,kumar2021some}.
\end{rem}
\subsection{Reductions under two-dimensional subalgebra}
The importance of two-dimensional subalgebra is that the reduced equation under one subalgebra will also be invariant under the second subalgebra and thus two-variable reduction can be achieved at one time. We have tried a reduction corresponding to each subalgebra in the list \eqref{BLP:10} and most can be discarded offhand as they are too trivial or due to the presence of arbitrary functions. Only the following possible reductions can be found that are significant.
\begin{case}
\normalfont Reduction under subalgebra $\mathcal{G}_{1}(X_{1}+\alpha\,X_{3},X_{2})$. The invariants of $X_{2}$ are 
\begin{align*}
    \xi=y, \eta=t, u=-\frac{x f_{2}^{\prime}(t)}{f_{2}(t)}+F(\xi,\eta), v=G(\xi,\eta).
\end{align*}
 The second generator in term of these invariants can be written in following manner,
\begin{align*}
    X_{1}+\alpha\,X_{3}=\alpha\,h_{1}(\xi)\frac{\partial}{\partial \xi}+4\,\frac{\partial}{\partial \eta}+0\,\frac{\partial}{\partial F}-\alpha\,\frac{d}{d\xi}(h_{1}(\xi))G\frac{\partial}{\partial G},
\end{align*}
where the arbitrary functions had to be fixed as $f_{1}(t)=1, f_{2}(t)=\mathrm{e}^{-2\,t}$. The corresponding auxiliary equations may written as,
\begin{align*}
    \frac{d\xi}{\alpha\,h_{1}(\xi)}=\frac{d\eta}{4}=\frac{dF}{0}=\frac{dG}{-\alpha\,h_{1}^{\prime}(\xi)}.
\end{align*}
Upon solving above auxiliary equations, a new set of similarity transformations for reduced equation under $X_{2}$ may be obtained as follows:
\begin{align*}
    \rho=\int\frac{1}{\alpha\,h_{1}(\xi)}d\xi-\frac{1}{4}\eta, F=\tilde{F}(\rho), G=\frac{1}{h_{1}(\xi)}\tilde{G}(\rho),
\end{align*}
and in term of $u$ and $v$, the above similarity transformations can be written as 
\begin{align}
 \label{BLP:14}   \rho=\int\frac{1}{\alpha\,h_{1}(y)}dy-\frac{1}{4}t, u=x+\tilde{F}(\rho), v=\frac{1}{h_{1}(y)}\tilde{G}(\rho).
\end{align}
The new similarity transformations \eqref{BLP:14} when substituted into system \eqref{BLP:1} will reduce it into a system of ODEs,
\begin{align*}
    \frac{d^{2}}{d\rho^{2}}(\tilde{F}(\rho))+8\,\frac{d}{d\rho}(\tilde{F}(\rho))=0, \frac{d}{d\rho}(\tilde{G}(\rho))=0. 
\end{align*}
The solution of above system shall provide a final solutions for the system \eqref{BLP:1}, that may be written as follow:
\begin{align}{\label{BLP:28}}
    u^{11}=c_{1}+x+\mathrm{exp}\left(-8\int\frac{1}{\alpha\,h_{1}(y)}dy+2\,t\right), v^{11}=\frac{c_{2}}{h_{1}(y)}.
\end{align}
\end{case}
\begin{case}
\normalfont Reduction under subalgebra $\mathcal{G}_{4}(X_{3}, X_{1})$. The invariance under this subalgebra tells that the solution of the system \eqref{BLP:1} must be of the form
\begin{align}
   \label{BLP:15} u=-\frac{xf_{1}^{\prime}(t)}{4f_{1}(t)}+\frac{1}{\sqrt{f_{1}(t)}}\tilde{F}(\rho), v=\frac{1}{h_{1}(y)}\tilde{G}(\rho), \rho=\frac{x^{2}}{f_{1}(t)}.
\end{align}
\end{case}
Substituting similarity transformations \eqref{BLP:15} back into the system \eqref{BLP:1}, we obtain following system of ODEs
\begin{subequations}{\label{BLP:16}}
\begin{align}
    &2\,\rho\frac{d^{3}}{d\rho^{3}}(\tilde{G}(\rho))+3\,\frac{d^{2}}{d\rho^{2}}(\tilde{G}(\rho))=0, \\
    &2\,\rho \frac{d^{2}}{d\rho^{2}}(\tilde{G}(\rho))+\frac{d}{d\rho}(\tilde{G}(\rho))+2\,\rho^{\frac{1}{2}}\tilde{F}\frac{d}{d\rho}(\tilde{G}(\rho))=0.
\end{align}
\end{subequations}
Immediate solution for the system \eqref{BLP:16}
\begin{align*}
    \tilde{F}(\rho)=\frac{c_{2}}{2\,c_{2}\rho^{\frac{1}{2}}+c_{3}}, \tilde{G}(\rho)=c_{1}-c_{2}\rho+c_{3}\rho^{\frac{1}{2}}.
\end{align*}
The final solution of the system \eqref{BLP:1} is obtained as follow:
\begin{subequations}{\label{BLP:29}}
\begin{align}
    &u^{12}=-\frac{xf_{1}^{\prime}(t)}{4f_{1}(t)}+\frac{c_{2}}{2\,c_{2}x+c_{3}\sqrt{f_{1}(t)}},\\
    &v^{12}=\frac{1}{h_{1}(y)}\left(c_{1}-c_{2}\frac{x^{2}}{f_{1}(t)}+\frac{c_{3}x}{\sqrt{f_{1}(t)}}\right).
\end{align}
\end{subequations}

\section{Conclusion}
The presence of Virasoro subalgebra in the infinite-dimensional Lie algebra \eqref{BLP:3} suggests that the Boiti–Leon–Pempinelli system \eqref{BLP:1} is completely integrable. To cover up the inadequacies in the earlier works \cite{yu2010exact,kumar2014new,kumar2015some,zhao2017lie,krishnakumar2020lie,kumar2021some},  a detailed groups classification of the infinite-dimensional Lie algebra \eqref{BLP:3} is obtained, and that too without assigning particular values to the arbitrary functions. One, two and three-dimensional groups classifications have been obtained at \eqref{BLP:8}, \eqref{BLP:10}, \eqref{BLP:11} respectively.

Further, a class of explicit solutions that involves several arbitrary functions of $y$ and $t$, is obtained in \eqref{BLP:18}--\eqref{BLP:27}. Using two-dimensional symmetry reduction, some more explicit solutions which involve arbitrary functions are also obtained in \eqref{BLP:28} and \eqref{BLP:29}.

\end{document}